\documentclass[%
reprint,
superscriptaddress,
preprintnumbers,
nofootinbib,
amsmath,amssymb,
aps,
prl
]{revtex4-1}

\usepackage{graphicx}
\usepackage{dcolumn}
\usepackage{bm}

\newcommand{\Exp}[1]{\operatorname{e}^{#1}}
\newcommand{\abs}[1]{\lvert{#1} \rvert}
\newcommand{\rmd}{{\mathrm{d}}}
\newcommand{\nn}{\nonumber}
\newcommand{\Lie}{\pounds}
\newcommand{\gLie}{\hat{\pounds}}
\newcommand{\rmT}{\mathrm{T}}

\newcommand{\AdS}[1]{\operatorname{AdS}_{#1}}
\newcommand{\rmS}{\operatorname{S}}
\newcommand{\OO}{\mathrm{O}}
\newcommand{\rmC}{\mathrm{C}}

\newcommand{\cE}{\mathcal E}
\newcommand{\cH}{\mathcal H}

\newcommand{\cS}{\mathcal S}

\newcommand{\ket}[1]{\lvert {#1} \rangle}
\newcommand{\bS}{{\mathring{\cS}}{}}

\newcommand{\CG}{G}
\newcommand{\OG}{\tilde{g}}
\newcommand{\OPhi}{\tilde{\phi}}

\newcommand{\bfr}{{\bm r}}
\newcommand{\bX}{{\bm X}}

\newcommand{\sla}[1]{\setbox0=\hbox{$#1$} 
\dimen0=\wd0 \setbox1=\hbox{/} \dimen1=\wd1 
\ifdim\dimen0>\dimen1 \rlap{\hbox to \dimen0{\hfil/\hfil}} #1 
\else\rlap{\hbox to \dimen1{\hfil$#1$\hfil}} / \fi}

\begin{document}

\preprint{KUNS-2670}

\title{Homogeneous Yang-Baxter deformations as generalized diffeomorphisms}

\author{Jun-ichi Sakamoto}
\email{sakajun@gauge.scphys.kyoto-u.ac.jp}
\affiliation{%
Department of Physics, Kyoto University, Kitashirakawa Oiwake-cho, Kyoto 606-8502, Japan}
\author{Yuho Sakatani}
\email{yuho@koto.kpu-m.ac.jp}
\affiliation{%
Department of Physics, Kyoto Prefectural University of Medicine, Kyoto 606-0823, Japan
}%
\affiliation{
Fields, Gravity \& Strings, Center for Theoretical Physics of the Universe, 
Institute for Basic Sciences, Daejeon 305-811, Korea
}%
\author{Kentaroh Yoshida}%
 \email{kyoshida@gauge.scphys.kyoto-u.ac.jp}
\affiliation{%
Department of Physics, Kyoto University, Kitashirakawa Oiwake-cho, Kyoto 606-8502, Japan}

\begin{abstract}
Yang-Baxter (YB) deformations of string sigma model provide deformed target spaces. We propose that homogeneous YB deformations always lead to a certain class of $\beta$-twisted backgrounds and represent the bosonic part of the supergravity fields in terms of the classical $r$-matrix associated with the YB deformation. We then show that various $\beta$-twisted backgrounds can be realized by considering generalized diffeomorphisms in the undeformed background. Our result extends the notable relation between the YB deformations and (non-commuting) TsT transformations. We also discuss more general deformations beyond the YB deformations. 
\end{abstract}

\maketitle

\setcounter{equation}{0}

{{\bf Introduction}.---}%
A fascinating topic in string theory is the AdS/CFT correspondence \cite{Maldacena:1997re}. 
The integrable structure plays an important role behind this gauge/gravity duality, 
and in order to find the extension away from the well-studied $\AdS5\times\rmS^5$ case, 
integrable deformations of the $\AdS5\times\rmS^5$ superstring have been eagerly studied. 
In particular, by employing the techniques of the \emph{Yang-Baxter (YB) deformations} 
\cite{Klimcik:2002zj,Klimcik:2008eq,Delduc:2013fga,Matsumoto:2015jja}, 
a framework to study a class of integrable deformations of $\AdS5\times\rmS^5$ superstring 
has been developed in \cite{Delduc:2013qra,Delduc:2014kha,Kawaguchi:2014qwa}. 
In this letter, we will concentrate on the \emph{homogeneous YB deformations}, 
which are based on classical $r$-matrices satisfying the homogeneous \emph{classical YB equation (CYBE)} 
\cite{Kawaguchi:2014fca,Matsumoto:2014nra,Matsumoto:2014gwa,Matsumoto:2015uja,vanTongeren:2015soa,vanTongeren:2015uha,Kyono:2016jqy,Osten:2016dvf,Hoare:2016hwh,Orlando:2016qqu,Borsato:2016ose,vanTongeren:2016eeb,Hoare:2016wsk,Borsato:2016pas,Hoare:2016wca,Araujo:2017jkb,Araujo:2017jap}.

Another intriguing topic in string theory is the $T$-duality. 
Especially, the familiar \emph{Abelian $T$-duality} \cite{Buscher:1987sk,Buscher:1987qj,Rocek:1991ps}, 
which is based on an Abelian isometry of a target space, 
has uncovered the hidden connection between string theories. 
On the other hand, its extension, so-called the \emph{non-Abelian $T$-duality} 
\cite{Fridling:1983ha,Fradkin:1984ai,delaOssa:1992vci,
Gasperini:1993nz,Giveon:1993ai,Alvarez:1994np,Elitzur:1994ri,Sfetsos:2010uq,Lozano:2011kb,Itsios:2013wd}, 
is still mysterious and needs further investigations. 
Recently, it was conjectured in \cite{Hoare:2016wsk} 
and proven in \cite{Borsato:2016pas} (see also \cite{Hoare:2016wca}) 
that a certain class of non-Abelian $T$-dualities are equivalent to the homogeneous YB deformations. 
It is then natural to expect that a deeper understanding of the homogeneous YB deformations 
will facilitate a further development of the non-Abelian $T$-dualities.

When we focus on the deformation of supergravity backgrounds, 
there are other useful techniques to obtain the YB-deformed backgrounds. 
The famous one is the \emph{TsT-transformation} \cite{Lunin:2005jy,Frolov:2005dj,Frolov:2005ty}, 
which is a combination of two (Abelian) $T$-dualities and a linear coordinate change (referred to as ``shift'').
As it has been noticed in 
\cite{Matsumoto:2014nra,Matsumoto:2014gwa,Matsumoto:2015uja,vanTongeren:2015soa,vanTongeren:2015uha,Kyono:2016jqy} 
and clearly shown in \cite{Osten:2016dvf}, all of the homogeneous YB deformations 
associated with Abelian $r$-matrices are equivalent to the TsT-transformations. 
A certain class of non-Abelian YB-deformed backgrounds can also be realized 
by a generalization of the TsT-transformation \cite{Orlando:2016qqu,Borsato:2016ose,vanTongeren:2016eeb}. 
In this letter, we develop this type of technique utilizing the framework of 
the \emph{double field theory (DFT)} 
\cite{Siegel:1993xq,Siegel:1993th,Siegel:1993bj,Hull:2009mi,Hull:2009zb,Hohm:2010jy,Hohm:2010pp,Jeon:2010rw,Hohm:2010xe,Jeon:2011cn,Hohm:2011zr,Hohm:2011dv,Jeon:2011vx,Hohm:2011si,Hohm:2011nu,Jeon:2011sq,Jeon:2012kd,Jeon:2012hp}, 
which provides a manifestly $T$-duality-covariant description 
for the massless sector of string theory.

As it has been observed in \cite{Arutyunov:2015qva}, 
some of YB-deformed backgrounds do not satisfy the usual supergravity equations 
but rather do the \emph{generalized supergravity equations (GSE)} \cite{Arutyunov:2015mqj,TW}. 
The DFT can reproduce both the usual and generalized supergravity from a single action 
\cite{Sakatani:2016fvh,Sakamoto:2017wor}, 
and it provides a unified description of YB-deformed backgrounds. 
The purpose of this letter is to elucidate that various YB-deformed backgrounds can be realized 
by performing a certain class of \emph{generalized diffeomorphisms} in the undeformed background. 
The generalized diffeomorphisms are the gauge symmetry of DFT, 
and the resulting deformed backgrounds automatically solve the equations of motion of DFT 
(as long as the consistency condition, called the \emph{strong constraint}, is satisfied). 
This ensures that the deformed background remains to be the string background.

In this letter, we consider YB-deformed backgrounds specified by classical $r$-matrices, 
$r=\frac{1}{2}\,r^{ij}\,T_i\wedge T_j$ ($\{T_i\}$: bosonic isometry generators of 
the undeformed background), satisfying the (homogeneous) CYBE,
\begin{align}
 f_{l_1l_2}{}^i\,r^{jl_1}\,r^{kl_2} + f_{l_1l_2}{}^j\,r^{kl_1}\,r^{il_2} 
 + f_{l_1l_2}{}^k\,r^{il_1}\,r^{jl_2} =0\,,
\label{eq:CYBE}
\end{align}
where $r^{ij}=r^{[ij]}$ is constant and specifies a deformation, and $f_{ij}{}^k$ 
is the structure constant $[T_i,\,T_j]=f_{ij}{}^k\,T_k$\,. 
As it has been noticed and proven in \cite{vanTongeren:2016eeb,Araujo:2017jkb,Araujo:2017jap} 
for the homogeneous YB deformations of $\AdS5$, the classical $r$-matrix has 
the interpretation as a non-commutative parameter in the dual open-string description. 
In terms of the generalized geometry \cite{Gualtieri:2003dx} or DFT, 
the correspondent of the non-commutative parameter is the $\beta$-field \cite{Duff:1989tf,Duff:1990hn} defined below. 
If we introduce the Killing vectors $e_i$ associated with the isometry $T_i$\,, 
the identification between the $r$-matrix and the $\beta$-field reads 
$\beta^{mn}\equiv r^{ij}\,e_i^m\,e_j^n$\,. 
This identification is true also for the homogeneous YB deformation of Minkowski space 
(see section 4.3 of \cite{Borowiec:2015wua} for an example that can be uplifted to ten dimensions), 
and we suppose that the identification works for arbitrary undeformed backgrounds. 
Our task here is to find out generalized diffeomorphisms which produce 
various $\beta$-twists specified by various $r$-matrices satisfying CYBE.

{{\bf A review of DFT}.---}%
In DFT, we consider a gravitational theory on a \emph{doubled space} with coordinates 
$(x^M)=(x^m,\,\tilde{x}_m)$ $(M=1,\dotsc,2D;\, m=1,\dotsc,D)$, 
where $x^m$ are the usual coordinates while $\tilde{x}_m$ are the dual coordinates. 
The generalized diffeomorphism in the doubled space is generated 
by the \emph{generalized Lie derivative},
\begin{align}
 \gLie_V W^M \equiv V^N\,\partial_N W^M - (\partial_N V^M-\partial^M V_N)\,W^N\,,
\end{align}
where the indices $M,N$ are raised or lowered with the $\OO(D,D)$ metric $(\eta_{MN})\equiv 
\Bigl(\begin{smallmatrix} 0 & \delta_m^n \\ \delta^m_n & 0 \end{smallmatrix}\Bigr)$. 
The generalized diffeomorphism is the gauge symmetry of DFT 
as long as the diffeomorphism parameter $V^M$ satisfies the \emph{weak constraint} 
$\partial_N \partial^N V^M=0$ and the strong constraint $\partial_N V^M\,\partial^N A=0$ 
where $A$ represents the parameter $V^M$ or the supergravity fields. 
The finite generalized diffeomorphism is realized by $\Exp{\gLie_V}$ \cite{Hohm:2012gk}. 
The gauge algebra $[\gLie_{V_1},\,\gLie_{V_2}]=\gLie_{[V_1,\,V_2]_\rmC}$ is governed 
by the \emph{C-bracket}, $[V_1,\,V_2]_\rmC 
\equiv \frac{1}{2}\,\bigl(\gLie_{V_1}V_2-\gLie_{V_2}V_1\bigr)$. 
For the usual vectors $V_a^M=(v_a^m,\,0)$ $(a=1,2)$ satisfying 
$\frac{\partial}{\partial \tilde{x}_m}V_a^N=0$, 
the C-bracket gives the Lie bracket; $[V_1,\,V_2]_\rmC=[v_1, v_2]$\,. 
In DFT, the bosonic fields consist of the \emph{generalized metric},
\begin{align}
 (\cH_{MN}) = \begin{pmatrix} \CG_{mn} -B_{mk}\,\CG^{kl}\,B_{ln} 
 & B_{mk}\,G^{kn}\\ -G^{mk}\,B_{kn} & G^{mn} \end{pmatrix}\,,
\end{align}
which integrates the (closed-string) metric and the Kalb-Ramond $B$-field, 
and the \emph{DFT dilaton} $d(x)$, which can be parameterized as 
$\Exp{-2d}=\sqrt{\abs{\CG}}\Exp{-2\Phi}$ ($\Phi$: usual dilaton), 
and an $\OO(D,D)$ spinor of the Ramond-Ramond (R-R) fields $\ket{A}$ 
(see \cite{Sakamoto:2017wor} for our conventions).

{{\bf The main result}.---}%
As discussed later, for a certain class of parameter $V_{(r)}$ specified by an $r$-matrix satisfying CYBE, 
the finitely transformed background is given by
\begin{align}
 &\cH^{(r)}_{MN}\equiv \Exp{\gLie_{V_{(r)}}} \cH_{MN} = \bigl(\Exp{\bfr^\rmT}\!\cH\Exp{\bfr}\bigr)_{MN}\,,
\nn\\
 &\bfr \equiv \begin{pmatrix} 0 & r^{mn} \\ 0 & 0 \end{pmatrix} \,, \quad d^{(r)}=d\,,\quad \check{F}^{(r)}=\check{F}\,,
\label{eq:beta-deformed-BG-DFT}
\end{align}
where $r^{mn}\equiv r^{ij}\,e_i^m\,e_j^n$\,. 
Suppose the absence of the $B$-field in the original background, 
then the deformed background in terms of the usual supergravity fields becomes
\begin{align}
\begin{split}
 &(\CG^{(r)}+B^{(r)})_{mn} = \bigl[(\CG^{-1}-r)^{-1}\bigr]_{mn} \,,
\\
 &\Exp{-2\Phi^{(r)}}=\Exp{-2 \Phi}\sqrt{\det[\delta_m^n-(\CG\,r\,\CG\,r)_m{}^n]} \,,
\\
 &\hat{F}^{(r)} = \Exp{-B^{(r)}_2\wedge}\Exp{-r\vee}\hat{F} \,,
\end{split}
\label{eq:beta-deformed-BG}
\end{align}
where $r\vee$ acts as $r\vee \hat{F} \equiv \frac{1}{2} \,r^{mn}\,\iota_m\iota_n \hat{F}$\,. 
As far as we know, all homogeneous YB-deformed backgrounds take the form 
\eqref{eq:beta-deformed-BG} irrespective of the original undeformed background. 
The formula \eqref{eq:beta-deformed-BG} suggests that 
the deformed background can be conveniently described by the dual fields, 
$(\OG_{mn},\,\beta^{mn},\,\OPhi)$ \cite{Andriot:2011uh}, defined through
\begin{align}
 (\OG^{-1}\!\! + \beta)^{mn}\! = \bigl[(\CG -B)^{-1}\bigr]^{mn}\,,\ \ 
 \sqrt{\abs{\OG}}\Exp{-2\OPhi}=\Exp{-2d} . 
\end{align}
In terms of the dual fields, the (open-string) metric $\OG^{(r)}_{mn}$ and the dual dilaton 
$\OPhi^{(r)}$ are the same as the original ones $\CG_{mn}$ and $\Phi$ 
while the $\beta$-field becomes $\beta_{(r)}^{mn}=r^{mn}$. 
In addition, a criterion whether the deformed background satisfies 
the usual supergravity equations of motion \cite{Borsato:2016ose}, 
called the \emph{unimodularity}, is expressed as
\begin{align}
 I^m \equiv D_n \beta_{(r)}^{mn} = - \frac{1}{2}\, r^{ij}\,[e_i, e_j]^m 
 \underset{\text{(unimodular)}}{=} 0 \,. 
\label{eq:unimodularity}
\end{align}
Here, the Killing property $D_m e_i^m=0$ is used and $D_m$ is the covariant derivative 
associated with the metric $\CG_{mn}=\OG^{(r)}_{mn}$. 
As we will discuss later, the formula \eqref{eq:beta-deformed-BG} is applicable 
for both the unimodular and non-unimodular cases, but in the latter case, 
the background follows the GSE with the extra vector $I^m$ given by \eqref{eq:unimodularity} 
as observed in \cite{Araujo:2017jkb,Araujo:2017jap}. 
It is also interesting to note that, in terms of the dual fields, 
CYBE \eqref{eq:CYBE} can also be expressed as
\begin{align}
 R \equiv [\beta_{(r)},\,\beta_{(r)}]_{\rmS} = 0\,,
\end{align}
where $[\ ,\ ]_{\rmS}$ denotes the Schouten bracket (see section 3 of \cite{Gualtieri:2003dx}). 
Namely, for the homogeneous YB-deformed backgrounds, 
the non-geometric $R$-flux \cite{Halmagyi:2009te} has to vanish.

{{\bf Almost Abelian twists}.---}%
Now, it is useful to explain the classification of the homogeneous YB deformations. 
An $r$-matrix $r =\frac{1}{2}\, r^{ij}\,T_i\wedge T_j$ is called \emph{Abelian} 
if it consists of a set of generators which commute with each other $[T_i,\,T_j]=0$, 
and otherwise called \emph{non-Abelian}. 
Most of homogeneous YB deformations studied in the literature are based on the Abelian $r$-matrices. 
The classification of non-Abelian $r$-matrix is very complicated in general, 
and it is mainly classified by the unimodularity condition \eqref{eq:unimodularity} 
(Abelian $r$-matrices are obviously unimodular). 
If we define the rank of an $r$-matrix as the number of generators contained in $r$, 
non-Abelian unimodular $r$-matrices with lower rank are classified well. 
Obviously, the rank-2 unimodular $r$-matrix is Abelian and the rank-4 unimodular $r$-matrix for the bosonic isometry of $\AdS5$ has been classified in \cite{Borsato:2016ose}. 
We here consider a class of unimodular $r$-matrices, called the \emph{almost Abelian} $r$-matrices \cite{vanTongeren:2016eeb}, 
which covers most of the rank-4 and rank-6 examples studied in \cite{Borsato:2016ose}. 
By using constant deformation parameters $\eta_i$ $(i=1,\dotsc,N)$, it takes the form, $r=r_N$ with
\begin{align}
 r_k \equiv \textstyle{\sum_{i=1}^k} \eta_i\,T_{2i-1} \wedge T_{2i}\,, \quad 
 [T_{2i-1},\, T_{2i}]=0\,, 
\label{eq:r_k}
\end{align}
and obviously satisfies the unimodularity condition \eqref{eq:unimodularity}. 
The \emph{almost Abelian} condition can be expressed as
\begin{align}
 [e_{2k-1},\,\beta_{(r_{k-1})}]_{\rmS} = 0\,,\quad [e_{2k},\,\beta_{(r_{k-1})}]_{\rmS}=0 \,,
\label{eq:almost-Abelian}
\end{align}
for $1\leq k\leq N$\,. 
This condition ensures CYBE. 
As argued in \cite{vanTongeren:2016eeb}, this class of YB deformations can be realized as a sequence of 
non-commuting TsT-transformations (see \cite{Borsato:2016ose} for the explicit form in the rank-4 examples), 
which consists of the usual TsT-transformations and diffeomorphisms that make the Killing vectors as coordinate basis. 
In principle, for a given almost Abelian $r$-matrix, it is possible to find non-commuting 
TsT-transformations and determine the resulting deformed background. 
However, it is a tough task in general, and in the following, 
by considering a specific type of generalized diffeomorphisms, 
we find the simple formula \eqref{eq:beta-deformed-BG} for $\beta$-twisted backgrounds associated with the almost Abelian $r$-matrix.

{{\bf Generalized diffeomorphism}.---}%
Let us introduce the generalized Killing vectors $E_i$ associated with $T_i$\,. 
For the usual isometries, $E_i$ take the form $(E_i^M)=(e_i^m,\,0)$ 
and satisfy $\eta_{MN}\,E_i^M E_j^N=0$. 
In addition, they are independent of the dual coordinates $\tilde{x}_i$\,. 
For an Abelian $r$-matrix, $r_1= \eta_1\,T_1\wedge T_2$, 
a coordinate system can always be found so that $e_2 = e_2^m\,\partial_m$ ($e_2^m$: constant) is realized. 
In such coordinates, we consider $V_{1}= \eta_1\, e_2^m\,\tilde{x}_m\,E_1$\,. 
Thanks to $[e_1, e_2]=0$ and the Killing property of $e_2$, 
$V_{1}$ satisfies the weak constraint $\partial_M\partial^M V_{1}=0$ and 
the strong constraint $\partial^M V_{1}^N\,\partial_M A=0$ 
where $A$ denotes $V_{1}$ or supergravity fields. 
Then, it is easy to show that 
\begin{align}
 \gLie_{V_{1}}\cH_{MN} = (\bfr^\rmT_1\,\cH + \cH\, \bfr_1)_{MN}\,, \ \
 \gLie_{V_{1}} d=0\,,
\label{eq:infinitesimal}
\end{align}
where $\bfr_1^{MN}\equiv 2\,\eta_1\,E^{[M}_1\, E^{N]}_2$\,. 
The R-R potentials $\check{C}$ and the field strengths $\check{F}$ are invariant. 
The finite transformation $\Exp{\gLie_{V_{1}}}$ gives \eqref{eq:beta-deformed-BG-DFT} with $\bfr$ replaced with $\bfr_1$. 
We then consider a further twist, $r_2=r_1+\eta_2\, T_3\wedge T_4$. 
From the almost Abelian property, we can again find a coordinate system where $e_4 = e_4^m\,\partial_m$ ($e_4^m$: constant) is realized, 
and perform a transformation $\Exp{\gLie_{V_{2}}}$ with $V_{2} = \eta_2\,e_4^m\,\tilde{x}_m\,E_3$\,. 
Repeating this procedure, we obtain the $\beta$-twisted background associated with the almost Abelian $r$-matrix, $r=r_N$\,.

In order to demonstrate the relation to the usual TsT-transformation, 
let us consider an Abelian $r$-matrix and choose a coordinate system where $e_i = \partial_i$ are realized. 
Then, our diffeomorphism parameter becomes 
$V = \sum_{i=1}^N\eta_i\, \tilde{x}_{2i}\, \partial_{2i-1}$ and it generates a generalized diffeomorphism, 
$x^M \to x'^M = \Exp{V}x^M$, or more explicitly, 
$x'^{2i-1} = x^{2i-1} + \eta_i \, \tilde{x}_{2i}$. 
This is nothing but the TsT-transformation in the DFT language.

As a non-trivial example, let us consider a deformation of $\AdS5\times \rmS^5$ background 
with the Poincar\'e metric,
\begin{align}
 \rmd s^2 =\frac{\rmd z^2 - 2\,\rmd x^+ \rmd x^- +(\rmd x^2)^2+(\rmd x^3)^2}{z^2} + \rmd s^2_{\rmS^5} \,. 
\end{align}
We denote the translation, Lorentz, and dilatation generators by $P_\mu$, $M_{\mu\nu}$, and $D$ ($\mu,\nu=+,-,2,3$), respectively, 
and consider a rank-4 $r$-matrix, $r=r_2$, with
\begin{align}
 T_1=M_{+2}, \ \, T_2=P_3,\ \, T_3=D-M_{+-}, \ \, T_4=P_+ .
\end{align}
These satisfy $[T_3,\, T_1] = T_1$ and $[T_3,\, T_2] = -T_2$, and constitute an almost Abelian $r$-matrix. 
This case, we consider a sequence of finite transformations 
$\Exp{\gLie_{V_2}}\Exp{\gLie_{V_1}}$ with
\begin{align}
 V_1 \equiv \eta_1\,\tilde{x}_3\,\hat{M}_{+2} \,,\quad 
 V_2 \equiv \eta_2\,\tilde{x}_+\, (\hat{D}-\hat{M}_{+-}) \,,
\end{align}
where hatted quantities like $\hat{M}_{+2}$ denote the generalized Killing vectors associated with the unhatted generators. 
These finite transformations produce the $\beta$-field,
\begin{align}
 \beta_{(r_2)}&=\eta_1\,\bigl(x^2\,\partial_+ + x^-\,\partial_2\bigr)\wedge \partial_3
\nn\\
 &\ + \eta_2\,\bigl(z\,\partial_z+2\,x^-\,\partial_-+x^2\,\partial_2+x^3\,\partial_3\bigr)\wedge\partial_+ \,,
\end{align}
and the deformed background \eqref{eq:beta-deformed-BG} is indeed a solution 
of type IIB supergravity. 
If one prefers to combine the transformations as a single one, 
$\Exp{\gLie_{V_2}}\Exp{\gLie_{V_1}}=\Exp{\gLie_{V_{12}}}$, 
the Baker-Campbell-Hausdorff formula \cite{Hohm:2012gk} would be useful.

{{\bf A more general class}.---}%
Let us consider a wider class of unimodular $r$-matrices, $r=r_N$ with \eqref{eq:r_k} satisfying
\begin{align}
 [e_{2k-1},\,\beta_{(r_{k-1})}]_{\rmS} = 0\,,\ \, 
 e_{2k-1}\wedge [e_{2k},\,\beta_{(r_{k-1})}]_{\rmS}=0,
\end{align}
for $1\leq k\leq N$, which covers all of the rank-4 unimodular $r$-matrices of $\AdS5$ \cite{Borsato:2016ose}, including the example where any TsT-like transformation has not been found. 
We explain a subtle issue in this class by considering the rank-4 example ($N=2$) 
where $[e_3,\,\beta_{(r_1)}]_{\rmS} = 0$ but $[e_4,\,\beta_{(r_1)}]_{\rmS} \neq 0$\,. 
Similar to the almost Abelian case, in coordinates where $e_2=\partial_2$, 
we first consider a finite transformation $\cH^{(1)}_{MN}=\Exp{\gLie_{V_1}}\cH_{MN}$ with $V_1=\eta_1\,\tilde{x}_2\,E_1$\,. 
Then, in coordinates where $e_4=\partial_4$, we perform the second transformation 
$\cH^{(2)}_{MN}=\Exp{\gLie_{V_2}}\cH^{(1)}_{MN}$ with $V_2=\eta_2\,\tilde{x}_4\,E_3$\,. 
According to $[e_4,\,\beta_{(r_1)}]_{\rmS} \neq 0$, $\cH^{(1)}_{MN}$ depends on the $x^4$ coordinate 
and hence the second transformation breaks the strong constraint; $\partial_K V_2\,\partial^K \cH^{(1)}_{MN} \neq 0$\,. 
In fact, the formula \eqref{eq:beta-deformed-BG-DFT} itself does not require the strong constraint, 
and indeed $\Exp{\gLie_{V_2}}\Exp{\gLie_{V_1}}$ provides the desired background. 
The problem is that a generic strong-constraint-violating generalized diffeomorphism is not a gauge symmetry of DFT. 
Therefore, the deformed background may not be a solution of DFT. 
Interestingly, for all examples in the list presented in \cite{Borsato:2016ose} 
(which cover all inequivalent rank-4 deformations of $\AdS5$), 
one can check that the equations of motion transform covariantly under the diffeomorphisms. 
At the present stage, we are not aware of the clear reason why such diffeomorphisms are allowed. 
A more general formulation of DFT 
\cite{Geissbuhler:2013uka,Blumenhagen:2014gva,Blumenhagen:2015zma}, 
where the strong constraint is rather relaxed, may help us to answer the question.

{{\bf Non-unimodular cases}.---}%
The last type of homogeneous YB deformations is the non-unimodular one. 
For simplicity, we will here focus upon the rank-2 Jordanian $r$-matrix, 
$r=\eta\,T_1\wedge T_2$ with $[T_1,\,T_2] = T_1$\,. 
In this case, the formula \eqref{eq:unimodularity} indicates that 
the unimodularity is broken: $I^m = -\eta\,e_1^m \neq 0$\,. 
For some non-unimodular cases, TsT-like transformations have been employed in \cite{Orlando:2016qqu} 
to reproduce the YB-deformed backgrounds on a case-by-case basis. 
Instead, we will here stick to our general strategy. 
In the present case, $[e_1,e_2]\neq 0$ introduces the $x^2$ dependence into $e_1$ 
and the parameter $V=\eta\,\tilde{x}_2\,E_1$ breaks even the weak constraint; $\partial_N\partial^N V^M\neq 0$. 
However, the formula \eqref{eq:beta-deformed-BG-DFT} still works 
due to the generalized Killing property of $E_1$ and the Jordanian property, $[e_1, e_2] = e_1$\,. 
A subtle issue is again the covariance of the equations of motion, 
and, as we show below, they are not transformed covariantly in the non-unimodular case.

In DFT, the generalized connection $\Gamma_{MNK}$ is supposed to transform as 
$\delta_V\Gamma_{MNK}=\gLie_V \Gamma_{MNK}-2\,\partial_M\partial_{[N}V_{K]}$\,. 
At the same time, it is defined to satisfy the condition 
$\nabla_M d\equiv \partial_M d+\frac{1}{2}\,\Gamma_K{}^K{}_M=0$\,. 
By the consistency, $\delta_V\nabla_M d=\gLie_V \nabla_M d + \partial^K V_M\,
\partial_K d - \frac{1}{2}\,\partial_N\partial^N V_M$ must vanish. 
It indeed vanishes if the strong constraint is satisfied. 
In the present case, the first two terms on the right-hand side vanish 
but the last term does not vanish because $V^M$ breaks the weak constraint. 
A short calculation shows $\delta_V\nabla^M d= \eta\,[E_1,\,E_2]_\rmC^M\equiv -\bX^M$. 
From the Jordanian property, the finite transformation gives $\Exp{\delta_V}\nabla^M d= -\bX^M$. 
Namely, after performing the deformation, 
$\nabla^M d$ does not vanish but becomes (minus) the null generalized Killing vector, 
$\bX^M=-\eta\,E_1^M=(I^m,\,0)$\,. 
This is the situation of the modified DFT (mDFT) \cite{Sakatani:2016fvh}, 
where the generalized connection is deformed by a null generalized Killing vector $\bX^M$. 
In the R-R sector, we suppose that the potential $\ket{A}$ and the field strength $\ket{F}$ 
transform covariantly; $\ket{A^{(r)}}=\Exp{\gLie_V} \ket{A}$ and $\ket{F^{(r)}} 
=\Exp{\gLie_V} \ket{F}$ (see \cite{Sakamoto:2017wor} for our conventions). 
However, the relation $\ket{F}=\sla{\partial}\ket{A}$ is deformed 
under the weak-constraint-violating generalized diffeomorphism as 
$\ket{F^{(r)}} =\sla{\partial}\ket{A^{(r)}} -\bX_M\,\gamma^M\,\ket{A^{(r)}}$, 
which is again the same relation as the one known in mDFT \cite{Sakamoto:2017wor}. 
The Bianchi identities (or the equations of motion) for the R-R fields are also deformed in a similar manner. 
It is tough to evaluate the deviation of the generalized Ricci tensors, 
$(\delta_V -\gLie_V)\, \cS_{MN}$ and $(\delta_V -\gLie_V)\, \cS$\,, 
under the weak-constraint-violating generalized diffeomorphisms. 
In fact, they do not vanish. For all of the rank-2 examples listed in \cite{Orlando:2016qqu}, 
we have checked that the following relations are satisfied:
\begin{align}
\bigl(\Exp{\bfr^\rmT}\!\cS\Exp{\bfr}\bigr)_{MN}=\bS^{(r)}_{MN}\,,\quad \cS=\bS^{(r)}\,. 
\label{eq:assumption}
\end{align}
Here, $\bS^{(r)}_{MN}$ and $\bS^{(r)}$ are modified generalized Ricci tensors 
\cite{Sakatani:2016fvh} in the deformed background. 
Then, since the stress-energy tensor obviously transforms covariantly, 
$\bigl(\Exp{\bfr^\rmT}\!\cE\Exp{\bfr}\bigr)_{MN}=\cE^{(r)}_{MN}$, 
the deformed background is a solution of mDFT. 
In fact, all solutions of mDFT can be mapped to solutions of DFT via a field redefinition \cite{Sakamoto:2017wor}, 
and the deformed background is still a solution of DFT.

To clearly see that the deformed background is indeed a solution of DFT, 
let us examine another route. For the example of $r=\eta\,P_-\wedge D$ \cite{Orlando:2016qqu}, 
instead of the weak-constraint-violating generalized diffeomorphism, 
we can find another generalized coordinate transformation 
which does not break the weak/strong constraint,
\begin{align}
\begin{split}
 z'&=(1 + \eta\,\tilde{x}_-)\,z\,,\quad x'^+ =(1 + \eta\,\tilde{x}_-)\,x^+\,,
\\
 \rho'&=(1 + \eta\,\tilde{x}_-)\,\rho\,,\quad \tilde{x}'_- = \eta^{-1}\,\log(1 + \eta\,\tilde{x}_-) \,.
\end{split}
\end{align}
Then, by employing Hohm and Zwiebach's finite transformation law \cite{Hohm:2012gk}, 
this transformation generates the same deformed background from the original $\AdS5\times \rmS^5$. 
In this route, instead of $\bX^M$, a linear $\tilde{x}_-$ dependence is introduced into the DFT-dilaton. 
This result is compatible with the (m)DFT picture discussed in \cite{Sakamoto:2017wor}.

{{\bf Twists by $\gamma$-fields}.---}%
So far, we have discussed only the homogeneous YB deformations, which always provide 
$\beta$-twists specified by the associated classical $r$-matrices. 
In the $U$-duality-covariant extension of DFT, called the \emph{exceptional field theory} (EFT) 
\cite{West:2001as,West:2003fc,West:2004st,Hillmann:2009ci,Berman:2010is,Berman:2011cg,Berman:2011jh,Berman:2012vc,Hohm:2013pua,Hohm:2013vpa,Hohm:2013uia,Aldazabal:2013via,Hohm:2014fxa}, 
we can consider a more general twisting via the $\gamma$-fields 
\cite{Aldazabal:2010ef,Chatzistavrakidis:2013jqa,Blair:2013gqa,Andriot:2014qla,Blair:2014zba,Sakatani:2014hba,Lee:2016qwn,Sakatani:2017nfr}. 
They are $p$-vectors dual to the R-R $p$-form potentials, and in particular, 
the bi-vector $\gamma^{mn}$ in type IIB theory is the $S$-dual of the $\beta^{mn}$ 
(see \cite{Sakatani:2017nfr} for the duality rules for these fields). 
In \cite{Kyono:2016jqy}, a YB-deformed background associated with $r=P_+\wedge (D-M_{+-})$ 
has been determined including the R-R fields, and indeed it is a solution of GSE \cite{Orlando:2016qqu}. 
Interestingly, a solution of standard supergravity 
which has the same NS-NS fields but different R-R fields has been found in \cite{Kawaguchi:2014fca}. 
In fact, the former is twisted by a $\beta$-field while the latter is twisted by a $\gamma$-field. 
At the supergravity level, the latter can be obtained by a combination of the TsT-transformations 
and the $S$-duality \cite{Matsumoto:2014ubv} (where the TsT-transformations generate 
a $\beta$-field and the last step converts the $\beta$-field to the $\gamma$-field). 
This deformation can also be realized as a generalized diffeomorphism in EFT. 
However, according to our current understanding, the YB deformation can produce only 
the former $\beta$-twisted background, and it is important to invent the extension, for example, 
by revealing YB deformations of the $S$-duality covariant $(p,q)$-string action \cite{Schwarz:1995dk}.

{{\bf Conclusion and Outlook}.---}%
In this letter, we have argued that homogeneous YB deformations 
associated with an $r$-matrix, $r=\frac{1}{2}\,r^{ij}\,T_i\wedge T_j$, 
can be regarded as a technique to provide $\beta$-twists with $\beta^{mn}=r^{ij}\,e_i^m\,e_j^n$\,. 
In this picture, the CYBE is equivalent to the absence of the $R$-flux. 
For undeformed backgrounds without the $B$-field, 
we have provided a simple formula \eqref{eq:beta-deformed-BG} for a general $\beta$-twisted background. 
Then, we have found the generalized diffeomorphism parameters which produce various $\beta$-twisted backgrounds, 
including all rank-4 deformations classified in \cite{Borsato:2016ose}. 
The advantage of our approach is that it 
(i) includes all the non-commuting TsT-transformations as special cases and 
(ii) can be applied to arbitrary undeformed backgrounds. 
In some examples, the associated diffeomorphism parameters break 
the strong or weak constraint, but the resulting deformed backgrounds remain to be solutions of DFT. 
Interestingly, for a general rank-2 Jordanian $r$-matrix, 
the generalized diffeomorphism has produced a non-unimodular deformation, 
and the extra vector $I^m$ appeared in the process of the diffeomorphism. 
We have also shown the novel relation, $I^m = D_n\beta_{(r)}^{mn}$, 
which is consistent with the result in \cite{Araujo:2017jkb,Araujo:2017jap}. 
To date, finite diffeomorphisms in DFT/EFT have not understood clearly 
\cite{Park:2013mpa,Hohm:2013bwa,Berman:2014jba,Cederwall:2014kxa,Papadopoulos:2014mxa,Hull:2014mxa,Cederwall:2014opa,Naseer:2015tia,Rey:2015mba,Chaemjumrus:2015vap,Bosque:2016fpi,Howe:2016ggg,Hassler:2016srl}. 
Hence we need to do more detailed analysis 
and clarify why the strong/weak-constraint-breaking diffeomorphism is allowed 
and how the global structure is deformed. 
It is also desirable to extend YB deformations and non-Abelian $T$-dualities. 
The duality covariant formulation, such as the \emph{double sigma model} and its extensions 
\cite{Tseytlin:1990nb,Tseytlin:1990va,Hull:2004in,HackettJones:2006bp,Hull:2006va,Copland:2011wx,Lee:2013hma,Blair:2013noa,Sakatani:2016sko,Park:2016sbw}, 
would be helpful along this direction. 

\medskip

{{\bf Acknowledgments}.---} 
The work of J.S. is supported by the Japan Society for the Promotion of Science (JSPS).
The work of K.Y. is supported by the Supporting Program for Interaction-based Initiative Team Studies 
(SPIRITS) from Kyoto University and by a JSPS Grant-in-Aid for Scientific Research (C) No.\,15K05051.
This work is also supported in part by the JSPS Japan-Russia Research Cooperative Program. 


\appendix

\section{Supplemental Material}

The Schouten bracket for a $p$-vector and a $q$-vector is defined by
\begin{align}
 &[a_1\wedge \cdots \wedge a_p,\,b_1\wedge \cdots \wedge b_q]_{\rmS} 
\nn\\
 &\equiv \sum_{i,j}(-1)^{i+j} [a_i,\,b_j] \wedge a_1\wedge \cdots \check{a_i}\cdots \wedge a_p
\nn\\
 &\qquad\qquad\qquad\qquad\qquad \wedge b_1\wedge \cdots \check{b_j}\cdots \wedge b_q \,,
\end{align}
where the check $\check{a_i}$ denotes the omission of $a_i$\,. 
In particular, for $p=1$, this coincides with the Lie derivative,
\begin{align}
 [a,\,b_1\wedge \cdots \wedge b_q]_{\rmS} = \Lie_a (b_1\wedge \cdots \wedge b_q)\,.
\end{align}

\end{document}